\documentclass[]{spie}  

\usepackage{amsmath,amsfonts,amssymb}
 \usepackage{wrapfig}
\usepackage{graphicx}
\usepackage{subcaption}
\usepackage[colorlinks=true, allcolors=blue]{hyperref}

\usepackage{aas_macros} 
\usepackage{xspace} 
\usepackage{upgreek} 
\newcommand{\um}{$\upmu$m\xspace}	
\newcommand{\textsim}{$\sim$} 

\title{Design Requirements for the Wide-field Infrared Transient Explorer (WINTER)}

\author[a,b]{Danielle Frostig} 
\author[c]{John W. Baker}
\author[e]{Joshua Brown}
\author[c]{Rick Burruss}
\author[e]{Kristin Clark}
\author[b]{Gábor Fűrész}
\author[c]{Nicolae Ganciu}
\author[b]{Erik Hinrichsen}
\author[d]{Viraj R. Karambelkar}
\author[d]{Mansi M. Kasliwal}
\author[b]{Nathan P. Lourie}
\author[b]{Andrew Malonis}
\author[a,b]{Robert A. Simcoe}
\author[c]{Jeffry Zolkower}

\affil[a]{MIT Department of Physics, 77 Massachusetts Ave., Cambridge, MA 02139, USA}
\affil[b]{MIT-Kavli Institute for Astrophysics, 77 Massachusetts Ave., Cambridge, MA, USA}

\affil[c]{Palomar Observatory, California Institute of Technology, Palomar Mountain, CA, USA 92060}
\affil[d]{Division of Physics, Math, and Astronomy, California Institute of Technology, 1200 E California Blvd, Pasadena, CA 91125, USA}
\affil[e]{MIT Lincoln Laboratories, Lexington, MA}
\authorinfo{Further author information: Send correspondence to D. Frostig at frostig@mit.edu. }

\begin{document} 
\maketitle

\begin{abstract}
The Wide-field Infrared Transient Explorer (WINTER) is a 1x1 degree infrared survey telescope under development at MIT and Caltech, and slated for commissioning at Palomar Observatory in 2021. WINTER is a seeing-limited infrared time-domain survey and has two main science goals: (1) the discovery of IR kilonovae and $r$-process materials from binary neutron star mergers and (2) the study of general IR transients, including supernovae, tidal disruption events, and transiting exoplanets around low mass stars. 

We plan to meet these science goals with technologies that are relatively new to astrophysical research: hybridized InGaAs sensors as an alternative to traditional, but expensive, HgCdTe arrays and an IR-optimized 1-meter COTS telescope. To mitigate risk, optimize development efforts, and ensure that WINTER meets its science objectives, we use model-based systems engineering (MBSE) techniques commonly featured in aerospace engineering projects. Even as ground-based instrumentation projects grow in complexity, they do not often have the budget for a full-time systems engineer. We present one example of systems engineering for the ground-based WINTER project, featuring software tools that allow students or staff to learn the fundamentals of MBSE and capture the results in a formalized software interface. We focus on the top-level science requirements with a detailed example of how the goal of detecting kilonovae flows down to WINTER’s optical design. In particular, we discuss new methods for tolerance simulations, eliminating stray light, and maximizing image quality of a fly’s-eye design that slices the telescope’s focus onto 6 non-buttable, IR detectors. We also include a discussion of safety constraints for a robotic telescope.

\end{abstract}

\keywords{Infrared Survey, Kilonova, Requirements Flowdown, Time domain survey}

\section{Introduction}




Over the past decade, there has been an rise in all-sky surveys and time-domain astronomy, especially in optical wavelengths\cite{rubin2019,ztf2019,PanStarrs2016}. Complementary infrared (IR) surveys can uniquely study dusty regions, cool stars, and possibly even detect transients powered by decay of $r$-process elements formed via neutron capture in binary neutron star mergers\cite{astro2020_ir}. Cost considerations, driven by sensors, have hitherto prevented parallel efforts to perform deep, cadenced surveys of the near infrared sky. The Wide-field Infrared Transient Explorer (WINTER) will deploy six 1920 x 1080 pixel InGaAs sensors to create a wide field-of-view mosaic in the SWIR waveband. Coupled with a 1-meter robotic telescope, WINTER will monitor the near infrared sky from Palomar observatory in the Y-band (1.0 \um), J-band (1.2 \um), and a shortened H-band (1.6 \um) down to a magnitude of $J_{AB} = 21$ in J. This paper is part of a series of papers describing WINTER, including a project overview (Ref.~\citenum{winter_overview}), a discussion of the readout electronics and InGaAs sensors (Ref.~\citenum{winter_sensors}), and a description of the fly's-eye lens mounting and optomechanics (Ref.~\citenum{winter_optics}). 

A primary interest of IR time-domain surveys is the search for kilonovae produced in the merger of two neutron stars, as is expected in the fourth observing (O4) run of the Laser Interferometer Gravitational-Wave Observatory (LIGO). Throughout O4, LIGO predicts the detection of 1-2 kilonovae per month \cite{2016LRR}. During nominal operations between gravitational wave events, WINTER will scan the accessible sky every two weeks to characterize the dynamic infrared sky. At WINTER's depth and waveband, transients of interest include: IR supernovae,  transiting exoplanets around dwarf stars, and tidal disruption events. In addition to transients, WINTER's survey will produce a deep co-added image of the IR sky to study galactic structure, high redshift QSOs, and Brown Dwarfs. Finally, WINTER will support general observing programs from the MIT and Caltech communities.  

WINTER’s design employs new technologies that require engineering and research efforts. It is the first large-scale IR astronomical program to utilize warm ($T\sim -50$ C) InGaAs sensors as an alternative to traditional, but expensive, HgCdTe. Accordingly, WINTER requires custom electronics, software, and cooling for reading the InGaAs CMOS focal plane arrays. To reformat the telescope's focal plane at high filling factor onto a set of six non-buttable sensors, we implement a novel field slicing design with 2:1 demagnification in each channel of the slicer, creating corresponding opto-mechanical mounting and alignment challenges. Finally, WINTER will operate autonomously, necessitating careful design of robotic software and safety systems.

WINTER is one example of the increasing complexity of modern ground-based instruments. We require careful planning to meet the science requirements with new technologies and a fixed budget. Similar to many modest, ground-based projects, WINTER does not have the resources for a full-time systems engineer. Instead, we present a small-group approach for creating a formalized science requirements flowdown with Model-based Systems Engineering (MBSE) techniques. MBSE manages complex systems from design through analysis, integration, and verification using system models. We implement our requirements flowdown using Systems Modeling Language (SysML) as implemented in NoMagic's \textit{MagicDraw} modelling tool. \textit{MagicDraw}'s GUI-based interface allows non-experts to learn the fundamentals of MBSE, with proper definition of requirement objects and their associated verification plan and rationale, and mapping of (often complex) dependency traces between levels. After the requirements are captured and organized, they are exported to HTML format and posted to our group web server so all members of the team can contribute to specific parts of the document and monitor requirements throughout the duration of the project. The software also stores data about the requirements in easily-interpreted json files that may be read into python scripts to simulate instrument performance as requirements are adjusted. In this approach, requirements can be monitored part time by a student or project staff or crowdsourced throughout the group.  


This paper presents the science to engineering flowdown for WINTER with an emphasis on the top level requirements (Section \ref{sec:top}), requirements for kilonovae detection (Section \ref{sec:kilo}), and robotic telescope safety (Section \ref{sec:safe}).

\section{Science Requirements Flowdown}

\label{sec:top} 
\begin{figure} [ht]
   \begin{center}
   \begin{tabular}{c} 
   \includegraphics[width=\textwidth]{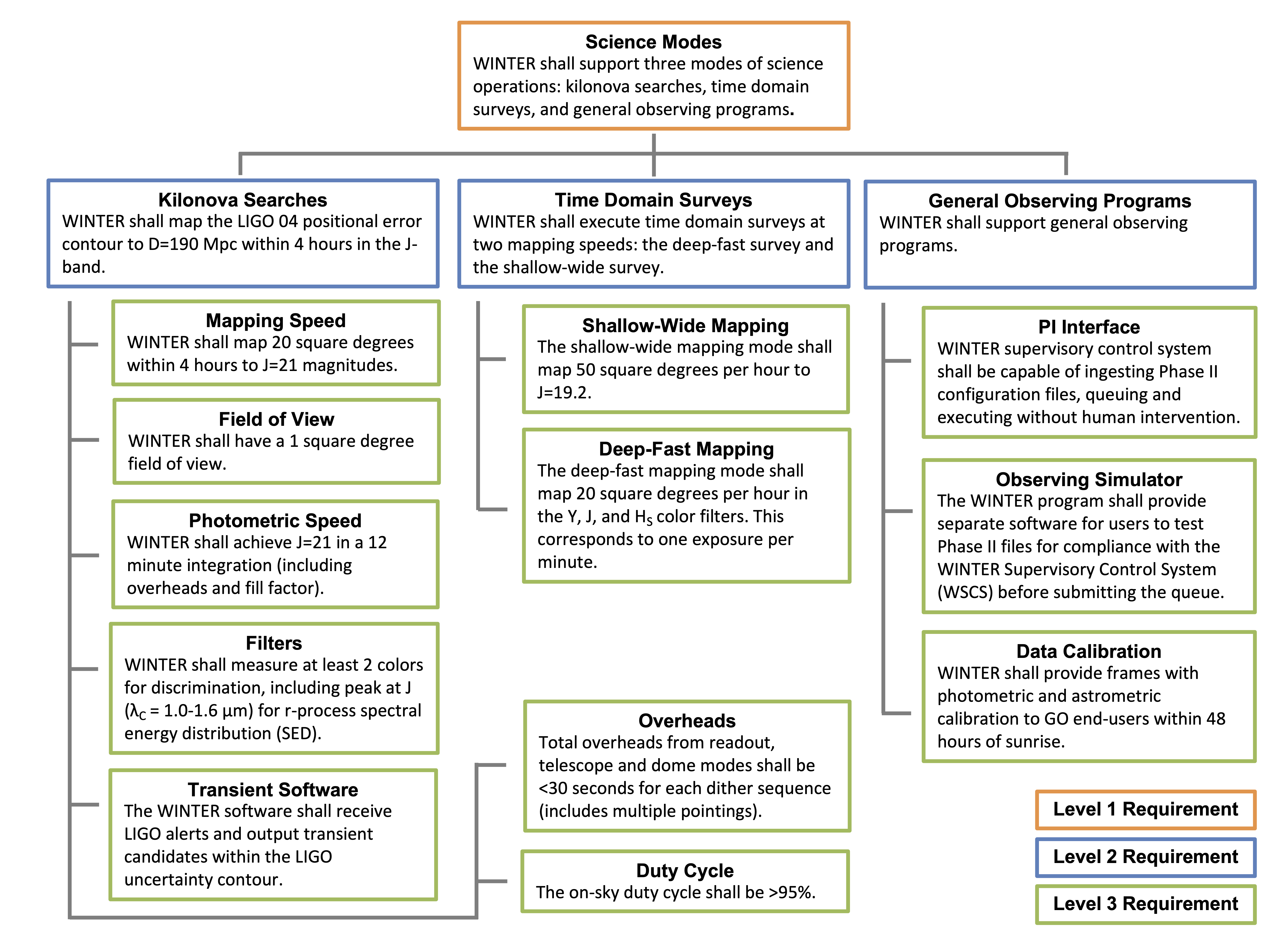}
   \end{tabular}
   \end{center}
   \caption[top] 
   { \label{fig:top} The top three levels of the WINTER science requirements flowdown tree. }
\end{figure} 

\begin{figure} [ht]
   \begin{center}
   \begin{tabular}{c} 
   \includegraphics[width=\textwidth]{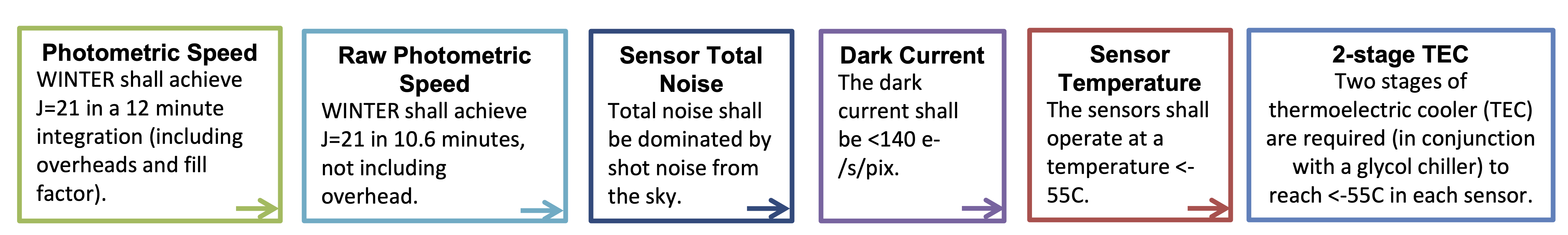}
   \end{tabular}
   \end{center}
   \caption[top] 
   { \label{fig:tec} An example requirement flowdown from the Level 3 photometric speed requirement to the necessity of a 2-stage thermoelectric cooler mounted behind each detector. }
\end{figure} 

For WINTER, we simplify the range of requirement types in MBSE to two broad categories: \textit{science requirements} are any requirements that flow down from the WINTER science modes and \textit{constraints} are non-science boundary conditions such as environmental history and safety protocols. Science requirements are discussed in this section and Section \ref{sec:kilo} and constraints are discussed in Section \ref{sec:safe}.

All science requirements flow down from the singular \textbf{Level 1 Requirement}--the Science Modes--with the explanatory text, “WINTER shall support three modes of science operations: kilonova searches, time domain surveys, and general observing programs” (Figure \ref{fig:top}). This requirement defines the general purpose of the project as a whole.

The \textbf{Level 2 Requirements} break the science modes into three programs with different technical specifications. The majority of WINTER requirements, such as image quality and mapping speed, flow down from the level 2 requirement for kilonova searches. When there are no LIGO detections for WINTER to observe, it will systematically conduct time domain surveys at two different cadences. The time domain surveys dictate requirements such as the faintest object WINTER will observe and the volume of data the project will produce and store. Finally, up to 25$\%$ of WINTER’s on-sky time will be devoted to the general observing program in which students and scientists can apply to use WINTER for their own projects. Requirements derived from the general observing program focus more on WINTER’s accessibility for general use and less on the unknown future science applications during general observing. 

The \textbf{Level 3+ Requirements} outline specific requirements that flow down from each observing program, from science modes down to specific engineering requirements. For example, Figure \ref{fig:tec} traces the level 3 photometric speed requirement derived for kilonova searches all the way down to the requirement to have 2 stages of thermoelectric cooler on each sensor. Laboratory tests running the sensors down to -50°C found that dark current scales exponentially with temperature and that the dark current halves with every additional 7°C of cooling \cite{Sullivan_2014}. WINTER’s science applications require that Poisson noise from the sky background exceeds sensor dark current and read noise, making detector cooling a priority for the instrument design. This requirement calls for a modification to the sensor housing and extra engineering activities. The example shown in Figure \ref{fig:tec} demonstrates the importance of defining all requirements early in project development to appropriately allocate engineering efforts across the project.

\section{Kilonova Search Requirements} \label{sec:kilo}

The majority of requirements for WINTER flow down from the science goal of mapping IR kilonova emissions from neutron star mergers. During its fourth observing run (O4), LIGO is expected to detect about 1-2 kilonovae per month out to a distance of 190 Mpc with a positional error contour of 20 square degrees \cite{2016LRR}. Assuming, on average, the region of uncertainty is above the horizon for 4 hours per night, we derive the Level 2 Requirement for Kilonova Searches: “WINTER shall map the LIGO O4 positional error contour to D=190 Mpc within 4 hours in the J-band.” The Level 3 Requirements further specify what is needed to achieve this goal:

\textit{Filters}:  The J-band specification in the Level 2 Requirement stems from two goals for WINTER: demonstrating near-IR InGaAs sensors can be employed in astronomical instrumentation and studying $r$-process emission from kilonovae. Models predict that during the merger of two neutron stars, heavy elements such as Lanthanides (atomic numbers 57-71) are synthesized through rapid neutron capture (the $r$-process). These short-lived, heavy nuclei then release energy via beta decay, powering an observable thermal transient known as a kilonova \cite{Li_1998}. Radiative transport calculations predict kilonova emissions with psuedo-blackbody spectra that peak and last longest in the Y (1.0 $\mu m$), J (1.2 $\mu m$), and H-bands (1.6 $\mu m$) of the near-IR \cite{Kasen_2013}. Observations of the kilonova GW170817 in gamma rays through radio waves led to more complex outflow models combining a short-lived polar jet peaking in blue wavelengths and an isotropic, longer-lived near-IR ejecta \cite{Kasen_2017}. 

\textit{Mapping Speed}: If GW170817, the first observed kilonova, occurred at the LIGO O4 observing horizon of  190 Mpc -- rather than the measured D\textless40 Mpc -- it would peak in the J-band at 21.0 magnitudes \cite{2016LRR}. Using this event as a limiting case, we derive the Mapping Speed requirement: “WINTER shall map 20 square degrees within 4 hours to J=21 magnitudes.” 

\textit{Field of View}: Mapping large portions of the sky quickly requires an instrument with the largest possible field of view (FoV). WINTER's field of view is approximately one square degree. 

\textit{Photometric Speed}: Splitting the four hours of dark time expected for each event across the twenty degree uncertainty contour allocates twelve minutes for each one square degree pointing. 

\textit{Duty Cycle}: In order to maximize mapping speed and cover the LIGO uncertainty contour in 4 hours, the on-sky duty cycle (the percentage of time WINTER is collecting photons) shall be greater than 95$\%$.

\textit{Transient Software}: The most efficient way to receive and map LIGO alerts is for WINTER to operate autonomously and therefore custom software must be written.  The WINTER data reduction pipeline will leverage heavily on past efforts undertaken for the Zwicky Transient Facility, and the Palomar Gattini-IR project \cite{Gattini_De_2020}.

\subsection{Image Quality}
\label{sec:image quality}
The WINTER science requirements flow down to over one hundred engineering specifications. As an example case, in this section we trace the requirement for kilonova searches down to design requirements for optical image quality, stray light mitigation in the camera, and optical alignment tolerances. 

\begin{figure} [ht]
   \begin{center}
   \begin{tabular}{c} 
   \includegraphics[width=\textwidth]{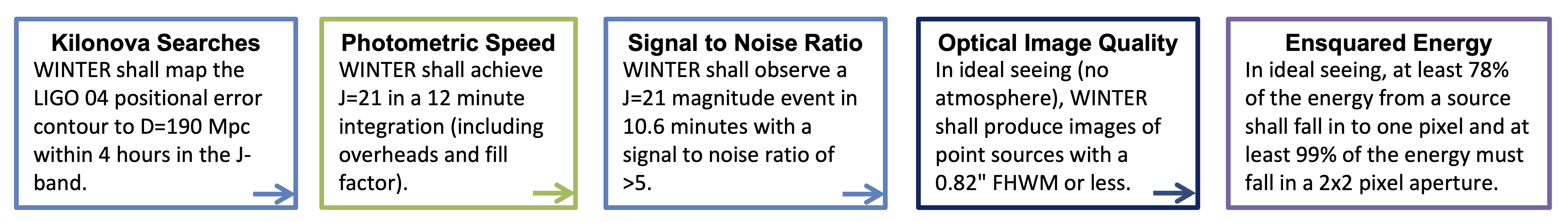}
   \end{tabular}
   \end{center}
   \caption[top] 
   { \label{fig:kilo} An example requirement flowdown from the Level 2 kilonova search to the required ensquared energy collected by each pixel. }
\end{figure}

\begin{figure}[t!] 

\begin{subfigure}{0.48\textwidth}
\includegraphics[width=\linewidth]{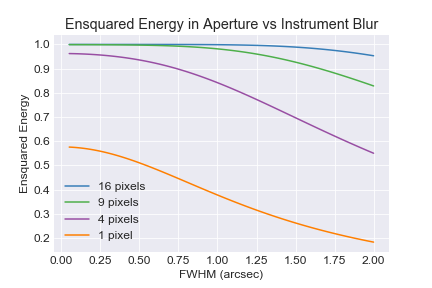}
\end{subfigure}
\begin{subfigure}{0.48\textwidth}
\includegraphics[width=\linewidth]{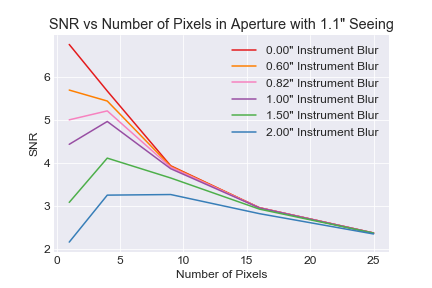}
\end{subfigure}\hspace*{\fill}
\caption{A study of ensquared energies and signal to noise ratios (SNR) for 1.1'' seeing at Palomar Observatory. The plot on the left compares the ensqaured energy in 1, 4, 9, and 16 pixel apertures for Gaussian beams of various widths. This in turn affects the SNR (right) for an instrument that adds 0.00'', 0.60'', 0.82'', 1.5'', and 2.0'' of blur. Each WINTER pixel is 15 \um x 15 \um. } \label{fig:study}
\end{figure}

For WINTER, we define a detection as an event with a signal to noise ratio (SNR) of 5 or greater. The SNR calculation (Equation \ref{eq}) includes noise levels and exposure time and therefore flows down from noise requirements and exposure time requirements that have already been specified and cannot be easily changed. We estimate Poisson noise from the sky from from Y-, J-, and H-band airglow measurements at Gemini Observatory \cite{10.1117/12.2233891} and set read noise and dark current upper bounds from based on a prototype instrument with InGaAs sensors \cite{Simcoe_2019}. 

The point source sensitivity follows the standard formula:

\begin{equation} \label{eq}
    SNR = \frac{\text{object counts}}{\sqrt{N_{Object} + (N_{Sky} + N_{Dark} + N_{Read \: Noise}^2)\cdot N_{pixels}}}
\end{equation}

Because sky, dark, and read noise all increase with the size of the photometric aperture, we are motivated to minimize instrumental broadening of the seeing disk. Ideally, with no blurring from the atmosphere or the instrument, a point source will fall onto one pixel and a J=21 magnitude event will be detected with an SNR of 14.8. Adding more pixels adds more Poisson noise from sky, dark current, and read noise, which collectively degrade the achieved SNR. The median site seeing at Palomar Observatory is 1.1” (arcseconds), which is similar to the size of a single WINTER pixel. WINTER's optical train will add more blurring; if varying amounts of blur are added in quadrature to the seeing disk. Figure \ref{fig:study} illustrates that we meet the requirement of point-source SNR=5 in a standard survey exposure for $J=21$ provided that this instrumental degradation is $<0.82$ arcseconds, and a $2\times 2$ pixel aperture is used. We meet these requirements with a five minute exposure time, adding a margin to the requirement to map one square degree with a ten minute expousre.


Having set requirements on the delivered FWHM, we calculated the equivalent ensquared energy for 1- and 4-pixel apertures on the sensor and input this to Zemax as our figure of merit for optical design optimization. For a beam with a FWHM of 0.82”, 78$\%$ of the ensquared energy (EE) from a source must fall in a 1 pixel aperture and 99$\%$ of the ensquared energy must fall into a 4 pixel aperture.  Achieving this specification over a 1-degree field requires careful optimization of both the optical design, and also a step-by-step strategy for adjusting optical compensators to take up errors in optical fabrication or optomechanical mounting.  At this level a well-aligned optical system would introduce a 20\% penalty to overall image quality. In practice our delivered design meets this requirement with comfortable margin.

\subsubsection{Optical Design}
\label{sec: optical design}
The image quality metrics described in the previous section formed the basis of an iterative design study, executed with Zemax OpticStudio\footnote{Zemax LLC, Kirkland, WA USA}, to optimize the WINTER optics design. The optimization merit function targeted a single-pixel geometric EE of 98\% across the field of view. This approach helps to balance the image quality across the full FoV, as opposed to simply maximizing the EE for each field which leads to higher throughput in the center of the field. Because the EE requirement was derived assuming a Gaussian point-spread-function (PSF), the merit function also targeted an RMS spot size \textless3 microns to act as loose constraint on the non-Gaussianity of the beam. 

Provisions for optical alignment, integration, and testing (AIT) were incorporated throughout the optimization and design process. Key constraints on the system layout were included in the optimization merit function to aid AIT and ensure that the camera could be adequately assembled separate from the telescope. The optimization enforced a minimum 2.5 mm external separation between the telescope focus and the first optical element to allow the instrument to be optically aligned using a pinhole mask. Using Zemax's multi-configuration functionality, the final round of optimization merit function included, for each waveband, image quality inputs for each  full system (instrument + telescope), as well as a laboratory (instrument only) configuration. This approach expands the optimization space for the full system (ie, by allowing telescope defocus to compensate for aberrations in the instrument optics) while ensuring the camera's standalone image quality is sufficient for aligning the instrument. 

\begin{figure} [ht]
   \begin{center}
   
   \includegraphics[width=\textwidth]{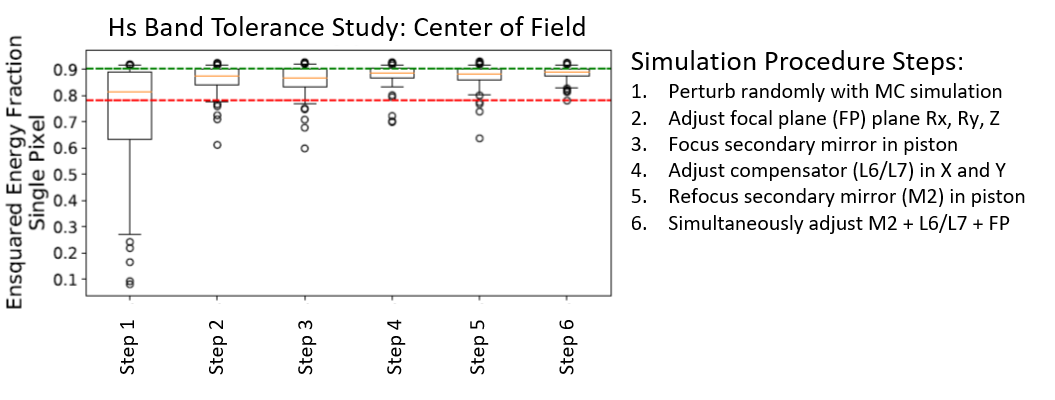}
   \end{center}
   \caption[Tolerance Study Example] 
   { \label{fig:tolerance} Example Monte Carlo (MC) tolerance study and AIT simulation, showing the single-pixel ensquared energy of each individual MC instance at each stage in the simulation of the AIT procedure. For each step, the box bounds the 25-75th percentile of the simulations, the whiskers bound the 5-95th percentile, and the individual points represent outliers. The green dashed line marks the nominal performance of the reference (unperturbed) optics design, and the red dashed line marks the 78\% minimum ensquared energy required to meet the observing requirements outlined in Section \ref{sec:image quality}.}
\end{figure}

The same integrated approach to incorporate AIT into the optical optimization guided the determination of the instrument fabrication and alignment tolerances. Using Zemax's Monte Carlo (MC) simulation capability, a detailed study was carried out to establish tolerances on the optical glass selection, lens fabrication, optomechanical support structures. Initial tolerances were determined from an inverse sensitivity analysis, which computed the minimum required change in a given parameter (ie, lens thickness, glass index of refraction, etc) for a specified change in the system merit function. These studies used a simplified merit function (as compared to the optimization merit function described in Section \ref{sec: optical design} which calculated the RMS energy not falling within a single pixel at six fields over the full FoV (ie: $\mathrm{MF_{Tol}} = 1-\sqrt{(1/N)\sum_{i=1}^6\mathrm{EE}_i^{\mathrm{pix}}}$. This sensitivity analysis also identified particularly sensitive degrees of freedom (DOF) in the design. The sensitivities were used to determine initial tolerances for each degree of freedom in the system and used as inputs to a series of AIT simulations. These simulations were automated using Zemax Programming Language (ZPL) scripts to simulate different alignment procedures to select the optimal approach. 

\subsubsection{Tolerancing}
 \begin{wrapfigure}{hrt}{0.47\textwidth}
  \centering
    \includegraphics[width=0.4\textwidth]{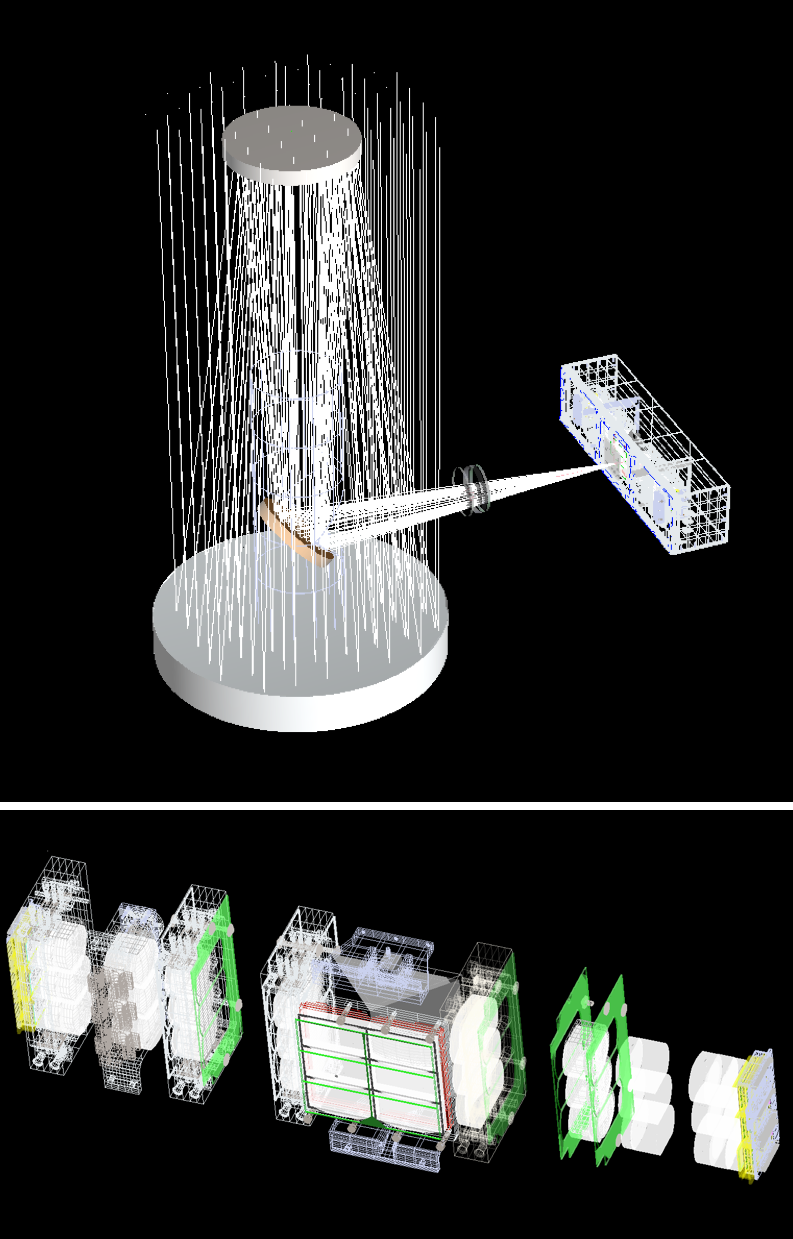}
  \caption{ A stray light analysis for WINTER. Top: Ray tracing a light source from the telescope optics to the instrument optics and mechanics (shown in the white box). Bottom: The instrument without its housing. The instrument is depicted with optomechanics on the left side of the instrument and no optomechanics towards the right for clarity. A few of the baffles in the system are shown in green.  }
  \label{fig:stray}
\end{wrapfigure}

The general approach of each simulation was to (1) perturb the reference optical design using Zemax's MC routine to randomly assign errors to each system DOF from a normal distribution with an RMS value determined from the earlier sensitivity analysis, (2) simulate optical adjustment by sequentially optimizing select optomechanical DOF to compensate for the degraded image quality, (3) add noise to the optimized motion of each compensator to simulate imperfect compensation in the lab. The degree of error added to the compensation reflected the capabilities of metrology equipment available for the WINTER integration. General coarse positioning of all optical components will be measured with a portable coordinate measuring machine (CMM) arm (10-25 \um accuracy); axial placement of each lens will be verified with a laser interferometer (\textless 1 \um accuracy). A typical result of one of these simulations is shown in Fig. \ref{fig:tolerance}. 

The final compensation approach selected for WINTER, based on the results of the tolerance studies, prioritizes a simplified alignment procedure at the cost of increased precision in the optics fabrication. By maintaining strict fabrication tolerances, and obtaining precise measurements of the glass indices over the full WINTER passband, the mechanical alignment tolerances were loosened to the scale of typical machining tolerances (\textsim25-50 \um). This approach requires a single compensating element which is precisely adjusted independently for each channel. The implementation of this approach is described in more detail in Refs. \citenum{winter_overview} and \citenum{winter_optics}.

\subsubsection{Stray Light and Filter Prescriptions}

To meet WINTER's sensitivity requirement, we must manage stray light in the form of either ghost images (from reflections off optical coatings), or scattered light (from lens edges, rays outside the clear aperture, or other opto-mechanical surfaces).  In particular, any nearly-in-focus image ghosts---whose position on the focal plane can move opposite to the direction of telescope pointing dithers---may be confused for a transient event and must be suppressed below detectable levels or handled in software. Our requirements specify that in-focus optical ghosts from a bright object should be attenuated by a factor of 10,000 relative to the primary or parent image of that object, equivalent to a 10-magnitude reduction.  Because WINTER will survey the full Northern sky to $J=21$, we will still detect ghost images in many fields with bright stars, and these will not be axisymmetric about the telescope boresight because of the separation of optical channels in the image slicer. Further attention has therefore been given to geometric mitigation of ghosts through intentional defocus of worst-offenders. We also plan for the data reduction pipeline to incorporate knowledge about how bright stars in different combinations of channel and field angle map into ghost images.

WINTER's stray light analysis uses the \textit{FRED} non-sequential ray tracing code from Photon Engineering, to model scattering and ghosts in the telescope and instrument optical train (Figure \ref{fig:stray}). As in many instruments, the most prominent stray light paths are formed from first reflection off the focal plane sensor, and then returned from back-reflection of optical coatings in the camera.  We mitigated these in WINTER's design by specifying efficient anti-reflection coatings on lenses near the detector, suppressing ghost images that originate from first-pass reflection off of the focal plane array.  We have also specified high performance coatings for the image slicing field lens near the telescope focal plane, which is another potential source of in-focus ghosts.

The FRED simulation also informed our strategy for placement of baffles, particularly those which prevent cross-talk between the six mosaic channels.  This cross-talk arises from scattered light in one optical train entering the clear aperture of the adjacent channel.  As seen in Figure 5, the telescope's image plane is sliced into six regions using a tiled ``fly's eye" field lens with a 3x2 array of rectangular apertures.  Our FRED model informed the design details and placement of a field stop baffle at the front end of the reimager, preventing light from reflecting off the bonded edges of the tiled lenses, while also maximizing the contiguous coverage of sky between sensors in the mosaic. Individual lens groupings in WINTER also feature sharp baffles to mitigate stray light. 

A final design edit that emerged from this process was relocation of WINTER's bandpass filters from the pupil located inside the reimager, to a position near the image slicer/field lens.  This change was motivated by the discovery that even for broadband filters, the wide field of WINTER's camera led to a field angle-dependent variation in the filter's central wavelength, because of the varying angle of incidence at the pupil.  WINTER's PlaneWave CDK1000 telescope has a highly telecentric focal plane, making it a more appropriate location for filters to eliminate field-dependent color terms in survey photometry.  As with the field lens, this presents a ghosting risk without proper attention to coating performance.  These surfaces are also displaced from the focal plane enough to generate a large amount of defocus for these images.

\section{Safety Constraints for Remote Operations} \label{sec:safe} 

\begin{figure} [ht]
   \begin{center}
   \begin{tabular}{c} 
   \includegraphics[width=0.72\textwidth]{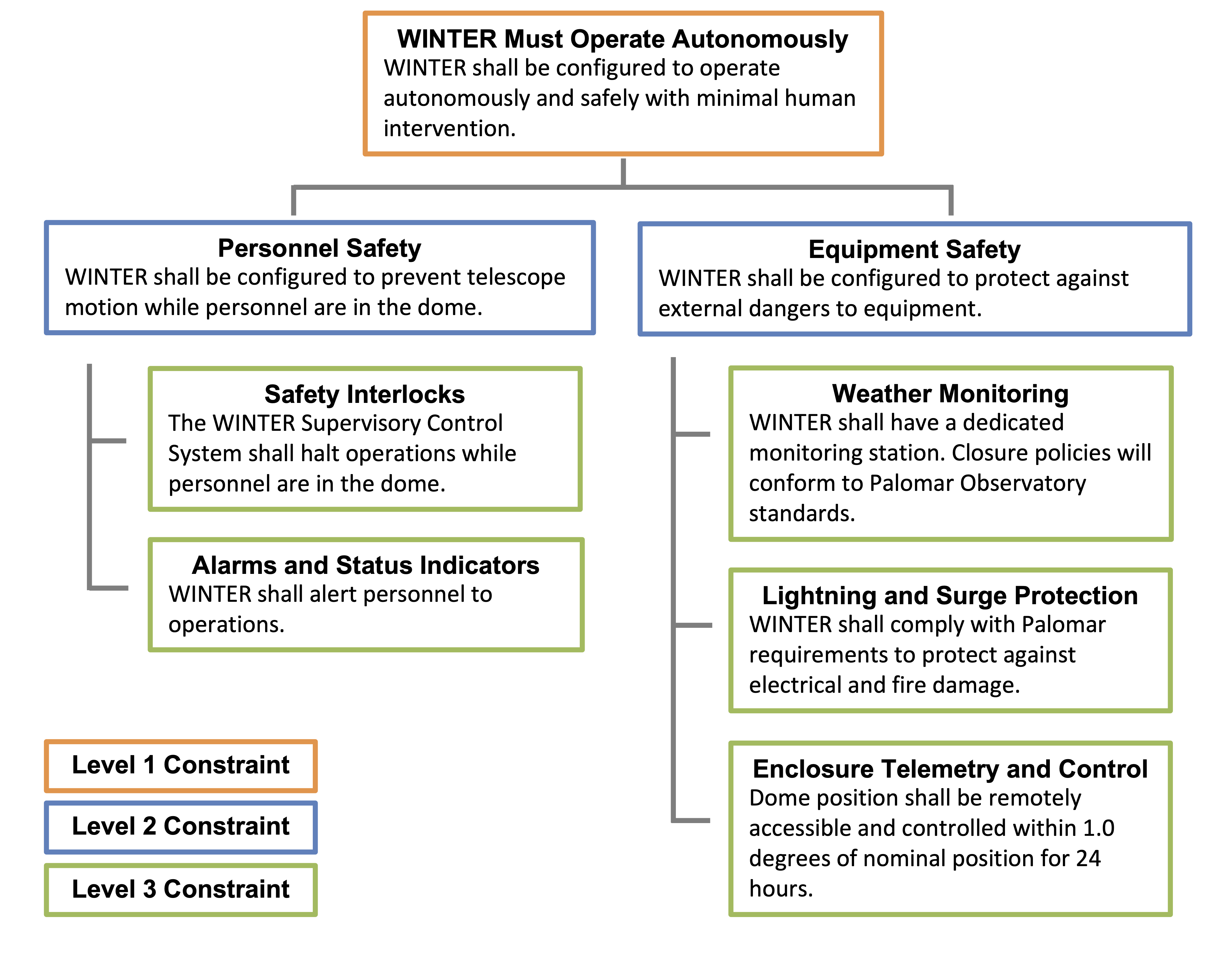}
   \end{tabular}
   \end{center}
   \caption[safety] 
   { \label{fig:safe} Constraints flowing down from the primary constraint of autonomous operations. }
\end{figure} 

For WINTER, constraints are defined as any necessary requirements that are not derived from the science modes, including environmental, safety, and technological constraints. Because WINTER does not have the budget for a telescope operator, it must operate autonomously (Figure \ref{fig:safe}). This is the primary constraint driving WINTER project requirements, as it flows into personnel and equipment safety and all subsequent requirements. To protect personnel, the telescope must halt slewing if anyone enters the dome. Safety interlocks will cease operations without human intervention, alongside a system of alarms and status indicators to alert personnel to telescope movement in case of an interlock failure. To protect the equipment, WINTER has a dedicated weather monitoring station, remotely accessible telemetry data for dome and telescope control, and fire protection measures.

\begin{figure} [ht]
   \begin{center}
   \begin{tabular}{c} 
   \includegraphics[width=0.66\textwidth]{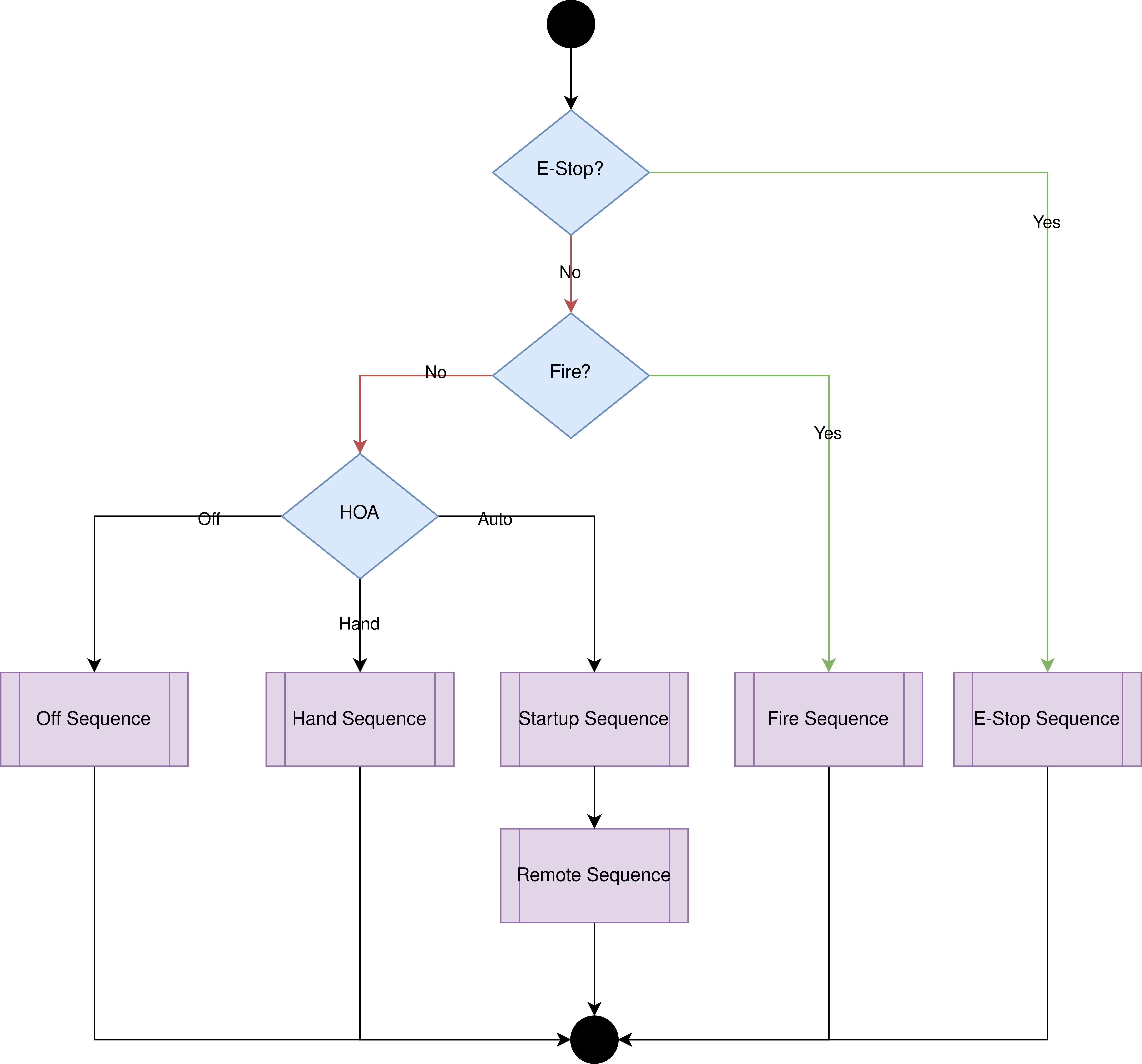}
   \end{tabular}
   \end{center}
   \caption[safety] 
   { \label{fig:overflow} A preliminary overview of the safety decision tree executed by the WINTER safety PLC (programmable logic controller). Each box represents a subsequent decision tree, with the remote sequence tree shown in Figure \ref{fig:remote}. }
\end{figure}

\begin{figure} [ht]
   \begin{center}
   \includegraphics[width=\textwidth]{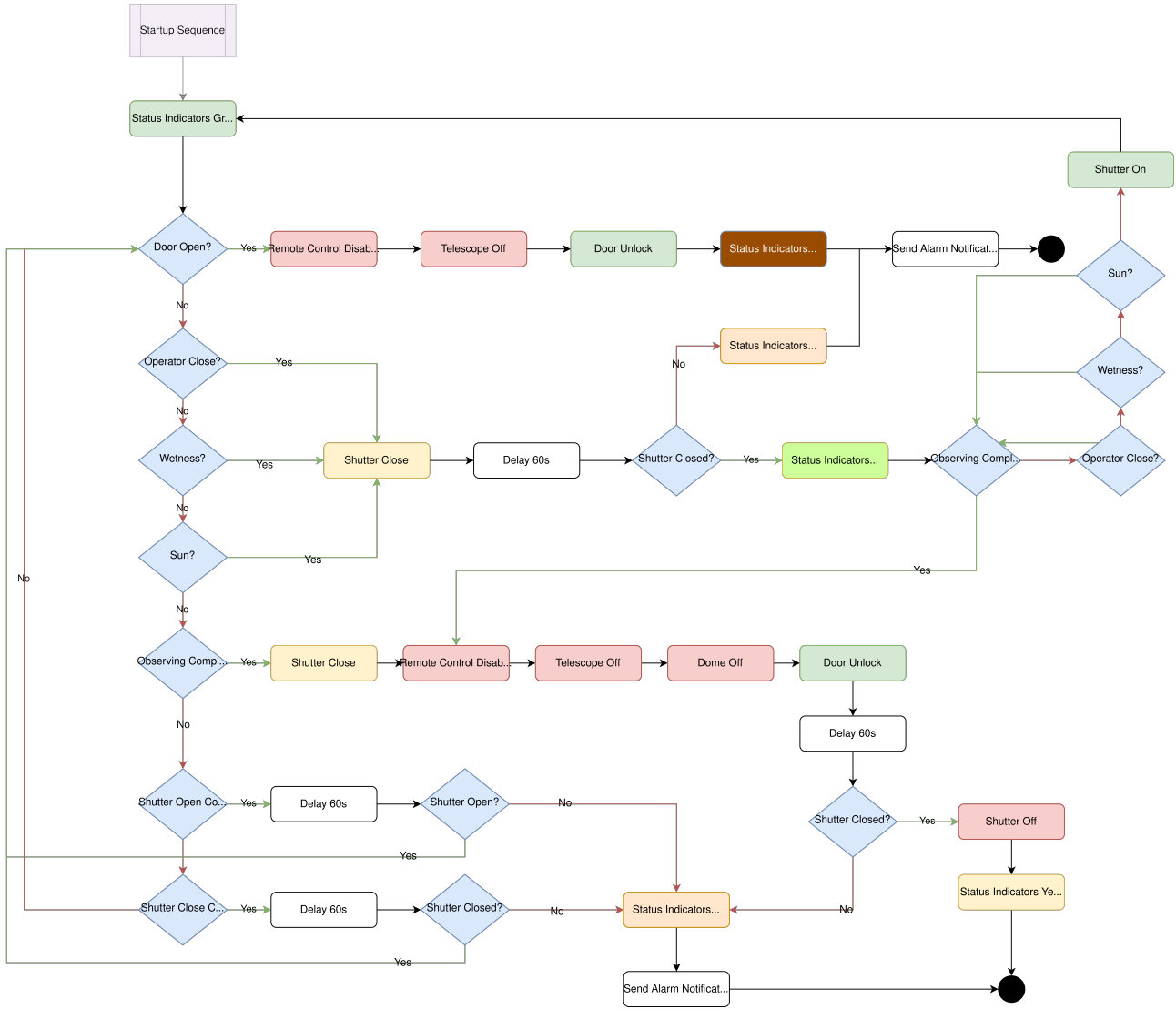}
   \end{center}
   \caption[safety] 
   { \label{fig:remote} A decision tree for remote observing with a robotic telescope. }
\end{figure}

The observatory has a series of interlocks which prevent startup (E-stop, alarms, etc) and initiate emergency sequences (Figure \ref{fig:overflow}). If no alarms or interlocks are tripped, the observatory state is dictated by a three-position hand/off/auto (HOA) key switch. This switch enables different operating modes. The off sequence safely shuts down the observatory, the hand sequence powers up all systems in an engineering/service mode which allows only local hand paddle control of the telescope, and the auto sequence initiates the nightly robotic control mode. The robotic control “remote sequence” is shown in detail in the diagram in Figure \ref{fig:remote}.

Other key constraints flow from current technology. A 1-meter telescope is the largest format commercially sold telescope and is the largest telescope that will fit in the available dome at Palomar Observatory. The 1920 x 1080 pixel InGaAs sensors are the largest on the market meeting our specifications, necessitating tiling 6 sensors together to achieve a wide enough field of view for an effective transient survey. Furthermore, the InGaAs substrate has a set quantum efficiency which affects the throughput of the instrument and therefore how long it needs to integrate at each pointing. Finally, conditions at Palomar Observatory, such as seeing and sky background noise, are constraints that influence the requirements for WINTER.

\section{Summary}

To meet specific science requirements, such as detecting kilonovae, setting clear requirements is critical to project success, especially when employing previously unexplored technologies in the field.  By employing tools from MBSE in a mid-scale ground based instrumentation project,  we can define and track key engineering tasks, appropriately allocate development efforts, and verify intermediate requirements throughout development. Placing these requirements on a team-accessible server in a simple, visual format also allows the entire team to monitor and update the requirements throughout a multi-year project.



\acknowledgments 

WINTER's construction is made possible by the National Science Foundation under MRI grant number AST-1828470.  We also acknowledge significant support from the California Institute of Technology, the Caltech Optical Observatories, the Bruno Rossi Fund of the MIT Kavli Institute for Astrophysics and Space Research, and the MIT Department of Physics and School of Science.

\bibliography{report} 
\bibliographystyle{spiebib} 

\end{document}